\documentclass{jpc} 


\usepackage{natbib}
\usepackage{amsthm}

\newtheorem{theorem}{Theorem}[section]
\newtheorem{definition}[theorem]{Definition}

\theoremstyle{plain}

\usepackage{multirow}
\usepackage{xcolor}

\usepackage{hyperref}
\hypersetup{
   colorlinks=true,
   citecolor=teal   
}

\usepackage[ruled, linesnumbered, noend]{algorithm2e} 
\usepackage{caption} 
\usepackage{subcaption}

\usepackage[textwidth=2.1cm]{todonotes}
\setuptodonotes{color=green!40}

\usepackage[skins, breakable]{tcolorbox}

\newtcolorbox{mybox}[2][]
{
  enhanced,
  rounded corners, 
  width = \columnwidth,
  colbacktitle=red!85!black,
  fonttitle=\ttfamily\small,
  fontupper=\small,
  attach boxed title to top left={xshift=2mm, yshift=-2mm},
  title={#2},#1
}

\begin{document}

\title[Protecting User Prompts Via Character-Level Differential Privacy]{Protecting User Prompts Via Character-Level Differential Privacy}


\author{Shashie Dilhara Batan Arachchige, Hassan Jameel Asghar, Benjamin Zi Hao Zhao, Dinusha Vatsalan, Dali Kaafar}	
\address{\centerline{Macquarie University, Cyber Security Hub, Sydney, Australia}}	





\begin{abstract}
  \noindent Large Language Models (LLMs) generate responses based on user prompts. Often, these prompts may contain highly sensitive information, including personally identifiable information (PII), which could be exposed to third parties hosting these models. In this work, we propose a new method to sanitize user prompts. Our mechanism uses the randomized response mechanism of differential privacy to randomly and independently perturb each character in a word. The perturbed text is then sent to a remote LLM, which first performs a prompt restoration and subsequently performs the intended downstream task. The idea is that the restoration will be able to reconstruct non-sensitive words even when they are perturbed due to cues from the context, as well as the fact that these words are often very common. On the other hand, perturbation would make reconstruction of sensitive words difficult because they are rare. We experimentally validate our method on two datasets, i2b2/UTHealth and Enron, using two LLMs: Llama-3.1 8B Instruct and GPT-4o mini. We also compare our approach with a word-level differentially private mechanism, and with a rule-based PII redaction baseline, using a unified privacy–utility evaluation.
  Our results show that sensitive PII tagged in these datasets are reconstructed at a rate close to the theoretical rate of reconstructing completely random words, whereas non-sensitive words are reconstructed at a much higher rate. Our method has the advantage that it can be applied without explicitly identifying sensitive pieces of information in the prompt, while showing a good privacy-utility tradeoff for downstream tasks. 

\end{abstract}

\maketitle

\section{Introduction}\label{Sec:Introduction}

With the rise of large language models (LLMs) many have started using them as helpful tools for personal and work-related tasks. These LLMs respond to prompts submitted by users to fulfill a given task. In many real-world scenarios, user prompts may include personally identifiable information, proprietary company data, or other confidential details. As these LLMs are mostly hosted by third parties, this risks leaking and exposing said information to third parties. For instance, in healthcare applications, LLMs might be prompted with clinical reports containing sensitive patient data to provide medical insights. Sensitive data in these reports may include patient names, medical record numbers, phone numbers, locations, and email addresses. These sensitive details could be memorized by the LLM, accidentally exposed in future interactions or even accessed by untrusted entities which can lead to severe ethical and regulatory violations~\citep{Tang2024}. Real-world reports show that such misuses of LLMs are not merely of theoretical concern. For instance, Samsung employees accidentally exposed confidential company meeting records and proprietary details of unreleased products by entering sensitive data into an LLM \citep{powell_samsung_2023}.

To address prompt privacy, various approaches have been proposed. These include sanitization using named entity recognition (NER) models and rule-based filtering~ \citep{Chen2024,kan2023protectinguserprivacyremote}, differential privacy (DP)~ \citep{Yue2021,Mattern2022,Tong2023,Du2023,Utpala2023,Feyisetan2019}, homomorphic encryption or secure multiparty computation (SMC) based techniques~\citep{curl-smc, llms-encrypted}, and custom techniques to obfuscate the prompt through for instance adding redundant or dummy tokens~\citep{Yao2024, Lin2024}, and simple encoding~\citep{Zhang2023}. 
Each of these techniques have their own limitations. 

NER and rule-based filtering methods rely on correctly defining and identifying sensitive entities in the text which may lead to false positives and negatives, as the NER model may miss some sensitive information while erroneously marking non-sensitive information as sensitive~\citep{10.1007/978-3-031-08473-7_30}.
Techniques based on homomorphic encryption and SMC~\citep{llms-encrypted, curl-smc} require changes to the LLM to enable computations in the encrypted domain, which in most circumstances is not realistic as these LLMs are hosted by third parties. Furthermore, currently, these techniques are slow due to significant computational and communication overhead. Custom obfuscation techniques~\citep{shen2024thiefkeeperbalancingusability, Zhang2023} do not carry formal privacy guarantees like homomorphic encryption and DP techniques. Furthermore, they also require architectural changes, where partial computation such as prompt tokenization is done at the client side. This again is not realistic in the setting where LLMs are hosted remotely by a third party. 

Differential privacy provides a mathematically rigorous framework for privacy of statistical databases~\citep{dwork2006calibrating}. However, applying it to the text domain is not straightforward, and has given rise to a number of different techniques. One such techniques is \emph{word-level} differential privacy~\citep{feyisetan2020privacy}. This replaces words in the text with other semantically similar words. However, this often impacts utility as it changes all words within the text which may or may not be sensitive. We provide a more detailed description and limitations of existing DP techniques for text sanitization in Section~\ref{sec:related-work}. Despite this, DP has one key advantage over the other techniques mentioned above: it can be applied in an automated manner without significant computational overhead as well as without needing access to any part of the remote LLM. 

In this paper, we propose a new way to apply differential privacy to sanitize text data given as input prompts to LLMs. Given a prompt to be sanitized, we perturb each character of every word of the prompt independently using a randomized response mechanism~\citep{warner1965randomized}. The resulting perturbed text is then restored by a remote LLM as much as possible and which subsequently performs the downstream task, all without access to the original, unperturbed prompt. 
The intuition is that \emph{prompt restoration} will reconstruct common words as each character in such words is correlated with other characters in the word, as well as other words in the prompt. On the other hand, sensitive words in the prompt tend to be rare and have higher entropy, and hence difficult to reconstruct by the remote LLM. This is backed up by previous results which show that LLMs handle typos and minor perturbations effectively for common words but struggle with rare words \citep{PromptRobust, Schick2020, Fox2022, Cao2023}. Since LLMs are the state-of-the-art for text-related tasks, if they are unable to reconstruct the original content, it is highly unlikely that any other method will perform better. The advantage of our mechanism is that we automatically provide privacy while maintaining utility of the prompt, without the need to explicitly identify or mark sensitive terms, using predefined rules or external classifiers, and thus avoiding the pitfalls of incomplete redaction, as may be the case with NER-based schemes.

\paragraph{In summary, our main contributions are as follows:}

\begin{itemize}
    \item We present a prompt sanitization method that applies local differential privacy at the character level using the $k$-ary randomized response ($k$-RR) mechanism directly on the user side, before the prompt is sent to an untrusted LLM provider. This reduces exposure of raw prompt content during transmission and removes the need for a separate sanitization component. It also removes dependence on NER or rule-based detection, since privacy protection is applied uniformly to the entire prompt.
    
    \item We leverage the capabilities of LLMs to reliably reconstruct common words over rare words, to ensure that the sanitized text remains both private and readable. We utilize two LLMs, GPT-4o mini and Llama-3.1 8B, to restore the LDP-sanitized text, allowing the LLM to recover many common words while minimizing the reconstruction of rare words, thereby balancing privacy and readability.

    \item 
    We establish a theoretical baseline for reconstruction of random words after they are perturbed by character-level $k$-RR. These words contain characters that are sampled uniformly at random, independently of other characters in the word. Using this baseline we show that sensitive PII is reconstructed at rates close to reconstruction of this baseline for random words; meaning recovery is no better than random guessing. In contrast, non-sensitive words are reconstructed at substantially higher rates, which supports the privacy and usability guarantees of our approach.

    \item 
    We benchmark our character-level $k$-RR method against a word-level differential privacy approach and rule-based PII redaction, under a unified privacy–utility evaluation. Privacy is quantified via sensitive term reconstruction rate (reported as privacy preserved), and utility is measured via cosine similarity between summaries from original versus protected inputs, under a standardized restoration and summarization pipeline. Our results indicate that character-level perturbation can achieve high summary similarity at higher levels of privacy preserved, compared to the word-level DP baseline. Presidio provides high utility at a single fixed setting, but lacks tunable privacy–utility trade-offs. To support reproducibility, we will release our code and evaluation scripts.

\end{itemize}

\section{Related Work}
\label{sec:related-work}
Prompt privacy comes under the umbrella of data sanitization methods which find and remove sensitive information, like personally identifiable information (PII), from data \citep{ishihara2023trainingdataextractionpretrained}.
Papadopoulou et al. \citep{52} introduce a neural privacy-enhanced entity recognizer for text sanitization, classifying personal identifiers. 
Kan et al. \citep{kan2023protectinguserprivacyremote} used a local LLM to sanitize privacy-sensitive user inputs before sending the sanitized text to a cloud-based LLM for processing. Instead of relying on a single local LLM for sanitization, Chen et al.\citep{Chen2024} introduced Hide and Seek (HaS), a framework with two models: the Hide-Model, which anonymizes private entities, and the Seek-Model, which restores them when needed. 

However, simple sanitization techniques are insufficient, as privacy depends on context and non-identifiable data \cite{40}. Furthermore, Carlini et al. \citep{24} demonstrate that sanitization is imperfect, often leaving private data exposed. Even with LLM-based sanitization, the level of protection depends on the model's ability to interpret context and strictly adhere to sensitivity instructions. Consequently, sensitive information may still be overlooked, particularly when contextual cues are subtle or the boundary of what is sensitive is ambiguous.

Local differential privacy (LDP) mechanisms provide a formal privacy guarantee by randomizing each user’s data on the client side, ensuring the true input is difficult to infer even for an untrusted observer \citep{duchi2013local}. In practice, LDP sanitizes sensitive data locally by introducing randomness into embedding vectors at the token, word, sentence, or prompt level before sending it to an untrusted LLM server \citep{Edemacu2024}.

Lyu et al. \citep{10.1145/3397271.3401260} proposed a local DP-based framework that ensures privacy by privatizing each word's representation vector and sequentially replacing sensitive words with semantically similar alternatives. To address privacy concerns related to previous differentially private mechanisms, Tong et al. \citep{tong2024inferdptprivacypreservinginferenceblackbox} introduced the InferDPT framework and proposed RANTEXT, a method that applies random adjacency for token-level perturbation in uploaded prompts. While differentially private mechanisms for perturbing word embeddings contribute to anonymization, \citep{mattern2022limitswordleveldifferential} highlight their limitations, including weakened theoretical privacy guarantees, vulnerability to deanonymization attacks, and challenges in preserving sentence content and language quality. Word-level privacy may fail to obscure higher-level concepts, leading Du et al. \citep{Du2023}  to explore sentence-level private embeddings, introducing metric LDP with two approaches based on Euclidean and angular distances. Utpala et al. \citep{Utpala2023} introduced DP-Prompt, a document-level DP method that uses zero-shot prompting on a local language model, applying a differentially private token selection process with the exponential mechanism to generate sanitized paraphrased documents for cloud LLMs.

Existing LDP text sanitization methods typically operate at the word, sentence, or document level. However, a major limitation of these approaches is that they indiscriminately perturb non-sensitive tokens alongside sensitive ones, often degrading the semantic coherence of the text and yielding a suboptimal privacy-utility tradeoff. In contrast, we apply LDP at the character level within the prompt and then use a remote LLM to recover utility by restoring many common words from context, while ensuring that sensitive, often rare, strings remain difficult to reconstruct.

Our design avoids key limitations of existing privacy-preserving techniques. Unlike NER-based sanitization \citep{Chen2024, kan2023protectinguserprivacyremote}, which depends on accurately detecting predefined entity types \citep{lison2021anonymisation, pilan2022text}, our method does not require entity detection. Under domain shift, novel formats, or obfuscation, NER can miss sensitive spans (false negatives), enabling verbatim leakage \citep{lison2021anonymisation, agarwal2021interpretability}. Furthermore, in contrast to cryptographic methods \citep{llms-encrypted, curl-smc} that incur high computational overhead, or heuristic obfuscation techniques \citep{Yao2024, Lin2024} that lack rigorous security definitions, our approach is computationally lightweight and offers formal privacy guarantees. Protection is applied uniformly at the character level, independent of whether the content matches any schema or requires changes to the remote model.

\section{Background}
\subsection{Large Language Models (LLMs)}
Language models (LMs) are computational models that are capable of understanding and generating human language and are able to predict the likelihood of word sequences or generate new text based on a given input \citep{49}. LMs can be down-streamed to various NLP tasks including Natural Language understanding and Generation, via transfer learning \citep{18}.  LMs can be Statistical Language Models (SLM), Neural Language Models (NLM), Pre-trained Language Models (PLMs) and  Large Language Models (LLMs) \citep{41}. \par
SLMs view text in a form of a word sequence and then estimate the probability of text as a product of their word probabilities \citep{41}. NLMs characterize the probability of word sequences by neural networks like Recurrent Neural Networks (RNNs) and use deep leaning methods to obtain knowledge of language patterns from  large text data corpora \citep{43}. Unlike NLMs, PLMs are domain independent and their training and inference, follows a pre-training or fine-tuning  approach \citep{42}. LLMs are advanced LMs, which are transformer-based \citep{46}  and contain a large number of parameters with exceptional learning capabilities of in context learning, instruction following, and multi-step following \citep{41} when compared to the PLMs.

\subsection{Differential Privacy}
\label{subsec:dp}
Differential privacy (DP) provides a mathematical framework for ensuring the confidentiality of an individual's data~\citep{dwork2006calibrating}. 
DP has been applied on two main settings, \emph{global} (GDP) and \emph{local} (LDP) \citep{GDPvsLDP}. GDP assumes the existence of a trusted data curator who knows the dataset in the clear, and applies a DP algorithm on the entire dataset. In the LDP setting each individual locally sanitizes his/her own data using a DP algorithm before passing on to the possibly untrusted curator~\citep{Kasiviswanathan}. Since our work applies to the local setting, this fleeting introduction to differential privacy is from the LDP lens.

Let $\mathcal{X}$ be the data domain, which we assume to be discrete. 

\begin{definition}[Local Differential Privacy]
\label{def:dp}
A randomized algorithm $\mathcal{M}: \mathcal{X} \rightarrow \mathcal{Y}$, with discrete range $\mathcal{Y}$, is $\epsilon$-differentially private if for all $x, x' \in \mathcal{X}$ and for all $y \in \mathcal{Y}$, we have:
\[
\Pr[\mathcal{M}(x) = y] \leq e^\epsilon \Pr[\mathcal{M}(x') = y],
\]
where $\epsilon \geq 0$. 
\end{definition}
In other words, anyone seeing the output is not able to discern the input except with the probability difference dictated by the parameter $\epsilon$. The following are known results.
\begin{theorem}[Post-processing Property~\citep{dwork2006calibrating}]\label{prop:post-processing}
    If $\mathcal{M}$ is $\epsilon$-DP, then for any algorithm $\mathcal{M}'$, $\mathcal{M}' \circ \mathcal{M}$ is also $\epsilon$-DP.
\end{theorem}
Note that $\mathcal{M}'$ does not take the initial data as input. It only access this indirectly through $\mathcal{M}$. 
\begin{theorem}[Sequential composition~\citep{dwork2014algorithmic}]\label{prop:composition}
    If $\mathcal{M}_1,\ldots,\mathcal{M}_t$ are $\epsilon$-DP, then the sequence of algorithms $\mathcal{M}' = \left(\mathcal{M}_1,\ldots,\mathcal{M}_t\right)$ is $t\epsilon$-DP.
\end{theorem}

\section{Proposed Method}
\subsection{Scenario and Threat Model}
Our scenario concerns a user who submits a prompt $T$ to an online remote LLM $\mathcal{R}$ hosted by a third party for some text generation task. This LLM is also our adversary. The prompt may contain sensitive terms which we do not want $\mathcal{R}$ to learn. The goal is to replace the prompt $T$ with a \emph{sanitized} prompt $T^*$ with the sensitive terms removed. This is done at the user side, e.g., using a Web browser plugin before the prompt is sent to $\mathcal{R}$.

\subsection{Overview of the Method and Intuition}
Our methodology consists of two main steps:

\begin{enumerate}
    \item \emph{Noise Injection.} We replace each character of every word in $T$ independently by a printable ASCII character via a local differentially private algorithm.   
    \item \emph{Prompt Restoration and Task Execution.} The noisy prompt is then given to a remote LLM $\mathcal{R}$, to restore as much of the text as possible by removing typos, and other errors, and perform the downstream task.
\end{enumerate}

\begin{figure*}[h]
    \centering
    \includegraphics[width=\linewidth]{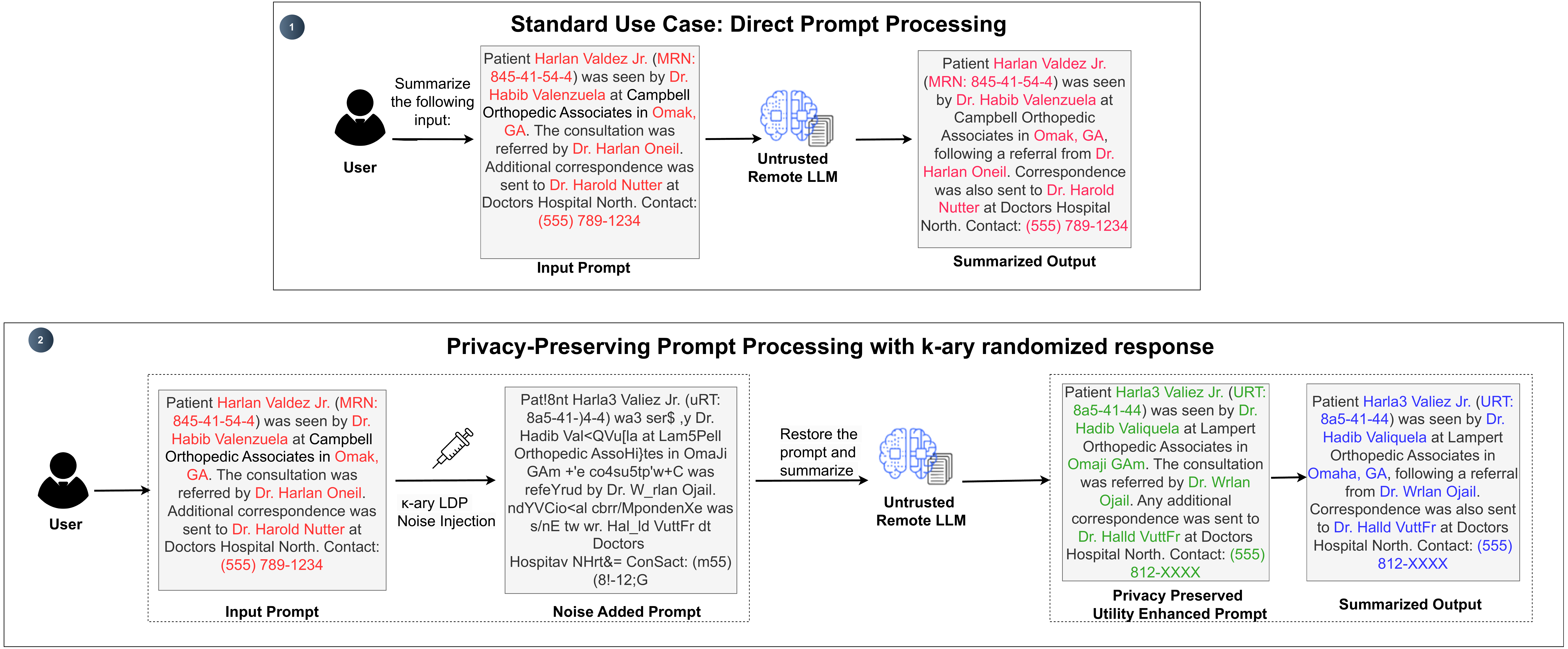}
    \caption{Proposed privacy-preserving prompt restoration pipeline with $k$-ary randomized response }
    \label{fig:Example_prompt}
    
\end{figure*}

After the first step, depending on the level of privacy via the privacy parameter $\epsilon$, some or all of the characters in every word in $T$ are replaced. However, LLMs can use their in-context learning capabilities to restore some or most of the original text. This is because common words in the prompt can be easily corrected by the remote LLM, even if some characters have been replaced, using cues such as the length of the word, its surrounding words and context. E.g., the perturbed word ``\verb+cbrr/MpondenXe+'' is correctly spell-corrected by the LLM as ``\verb+correspondence+'' in Figure~\ref{fig:Example_prompt}. On the other hand, many sensitive terms are rare, and hence the LLM is unable to reconstruct them correctly, e.g., \verb+Harlan+ $\rightarrow$ \verb+W_rlan+ $\rightarrow$ \verb+Halld+, again in Figure~\ref{fig:Example_prompt}. This simple observation leads to an automated way of sanitizing the prompt while maintaining utility. 

For the purpose of our study, we conceptually separate the overall process handled by the remote LLM $\mathcal{R}$ into two sequential stages: \emph{prompt restoration} and \emph{task execution}. Although both stages are performed by $\mathcal{R}$ in practice as part of a single inference step, this conceptual separation allows us to analyze how the model first attempts to restore the noised input and subsequently performs the intended downstream task. This decomposition provides a clearer understanding of the remote LLM's behavior in the presence of privacy-preserving perturbations.

\subsection{Defining Sensitive Terms}
It is difficult to come up with an all encompassing definition of privacy for user prompts as it differs depending on the case. For instance, one can think of scenarios where the entire prompt may be considered sensitive. We therefore do not claim to provide privacy for every use case. Rather, we focus on a scenario which is arguably commonplace:~\citep{gomez2010data, dernoncourt2017patientnotesner, joshi2022ner} we assume that there are only a certain categories of words in the text which are flagged as sensitive such as full names of individuals, email addresses, and residential addresses. These are commonly known as personal identifiable information (PII). Furthermore, these categories may be dataset dependent. See Section~\ref{subsec:datasets} for the categories used in our experiments.

Returning to our discussion in the previous section, the words belonging to these categories, which we call \emph{sensitive words}, are often \emph{rare}, in the sense that they are either never or rarely found in text corpuses, such as a private residential phone number. Since LLMs are generally trained on text data available from the Web, these words would be rarely encountered by them, and hence would be difficult for them to reconstruct if distorted via our method. On the other hand, many instances of these words are indeed common, e.g., Barack Obama. These are more likely to be reconstructed by LLMs as they have seen them frequently in their training data. However, because these terms are widely known and publicly available, their reconstruction poses a lower privacy risk. Thus, our method provides an automated way of striking a balance between privacy and utility: rare PII remain effectively anonymized, while common terms are handled in a way that preserves contextual coherence.

\subsection{Noise Injection via $k$-Randomized Response} \label{kRR_algo}
Let $C$ denote the set of printable ASCII characters minus the `space' character. Thus $|C| = 94$. A word $w$ is a sequence of characters from $C$. The $i$th character in $w$ is denoted $w[i]$. Let $|w|$ denote the length of the word $w$. For any two words $w, w'$ of equal length let $d(w, w')$ denote their Hamming distance which counts the number of characters different in the two words. Thus, $0 \leq d(w, w') \leq |w| = |w'|$.
We employ the $k$-Randomized Response ($k$-RR) algorithm~\citep{warner1965randomized, balle2019shuffle} described in Algorithm~\ref{algo:krr} to perturb each character of a word. For a single character, the $k$-RR is $\epsilon$-differentially private. For completeness, we provide a proof in Appendix~\ref{app:krr-proof}. Using the sequential composition theorem (Theorem~\ref{prop:composition}), the mechanism is $|w|\epsilon$-differentially private for a word $w$ of length $w$. 

\begin{algorithm}
\SetAlgoLined
\DontPrintSemicolon
\SetKwInOut{Input}{Input}
\Input{Character set $C$, word $w$, privacy parameter $\epsilon$}
Set $k \leftarrow |C|$\;
Set $\gamma \leftarrow \frac{k-1}{k - 1 + e^\epsilon}$\;
Initialize $w^* \leftarrow \emptyset$\;
\For{$i = 1$ \KwTo $|w|$}{
    $b \leftarrow \text{Ber}(\gamma)$\;
    \eIf{$b = 0$}{
        $w^*[i] \leftarrow w[i]$\;
    }{
        $w^*[i] \overset{{\scriptscriptstyle\$}}{\leftarrow} C - \{w[i]\}$\;
    }
}
\KwRet{$w^*$}
\caption{$k$-Randomized Response~\citep{warner1965randomized, balle2019shuffle}}
\label{algo:krr}
\end{algorithm}

Given some the input prompt $T$, the noise injection step applies the $k$-RR algorithm to each word in $T$, which is then passed on to the remote LLM $\mathcal{R}$ to restore the prompt and perform the downstream task. Prompt restoration and subsequent task remains differentially private due to the post-processing property of differential privacy (Theorem~\ref{prop:post-processing}). We give details on how this LLM is instructed to restore the prompt and perform the subsequent task in Section~\ref{subsec:prompt-restore}. We denote the prompt after the noise injection by $T^*$. Note that the $k$-RR algorithm maintains the length of words. The length of words is also \emph{mostly} preserved after the post-processing step by the LLM. Thus, we assume a one-to-one correspondence between the position of each character in the original word $w$ in $T$, and the transformed word $w^*$ in $T^*$.

\subsection{Privacy Interpretation}
In Section~\ref{subsec:dp}, we defined DP in terms of hiding an individual's data. Our use case deviates from this typical DP scenario, and therefore  privacy guarantees need to be interpreted differently. We apply LDP to all characters of a word in the prompt. Privacy is breached if the perturbed word can be reconstructed in the post-processing step. The most appropriate way to understand the privacy guarantee provided by our mechanism is to think of correlations between characters of a word. If a word is completely random, i.e., each character is sampled uniformly at random from $C$, then there is no correlation between characters. In this case, the differentially private mechanism ensures that the probability of reconstructing the original word is only slightly better than randomly guessing each of its characters, depending of course, on the choice of $\epsilon$. 

On the other hand, if a word is common in a language, say the English language, then not only are the characters of this word correlated with other characters of the word, but also with other words in the prompt. It has previously been flagged that one needs to carefully examine privacy expectations if the input data to a differentially private algorithm is strongly correlated~\citep{kifer2011freelunch}. For instance, a family of 10 living under one roof is either all susceptible to a disease or none of them are, even if each individual submits a (binary) randomized response to the question of whether he/she has the disease~\citep{kifer2011freelunch}. Thus, the privacy guarantee in this case is weaker. But this is precisely what we want! 

In many use cases of data sanitization, common words are rarely sensitive, and therefore weaker privacy protection for them is not only tolerable, but also desirable from a utility point of view. On the other hand, sensitive words are often rare, and exhibit properties similar to the uncorrelated setting. 

As we further discuss in Section~\ref{subsec:Reconstructed_Words}, unlike common words, rare words do not follow strong correlations or statistical dependencies, making them harder for LLMs to reconstruct based on fewer contextual clues available.
In the following, we analyze how difficult it is to restore words with uncorrelated characters after being passed through the $k$-RR algorithm, which establishes our baseline.

\subsection{Word Reconstruction Analysis and Baseline}
\label{sec:krr-analysis}
Given an arbitrary target character of some word $w^* \in T^*$, we are interested in knowing how likely it is for any algorithm to guess the character in the same position of the original word $w$. 

To estimate this, let us define some events. Let $\mathsf{C}_i$ be the event that the target character in $w$ is the $i$th character from $C$. The probability space of $\mathsf{C}_i$ is over the distribution of characters and words in the corpus. Let $\mathsf{C}^*_i$ be the event that the target character in $w^*$ is the $i$th character from $C$. The probability space of $\mathsf{C}^*_i$ is, in addition to the aforementioned distribution, also over the randomization in the $k$-RR algorithm. Likewise let $\mathsf{W}$ be the event that the target word is $w$, and $\mathsf{W}^*$ be the event that the word $w^*$ is output by $k$-RR. 
For $0 \leq j \leq |w|$, let $\mathsf{W}_j$ be the event that a word $w'$ is within Hamming distance $j$ from the target word $w$, i.e., $d(w, w') \leq j$. Note that $\mathsf{W}_0 = \mathsf{W}$. 

Given the description of $k$-RR, privacy parameter $\epsilon$, the vocabulary, and $T^*$, the goal of the algorithm is to correctly guess the target character in $w$. To estimate the success probability of the algorithm, we calculate the probability $\Pr[ \mathsf{C}_i \mid \mathsf{C}^*_j]$, i.e., the probability that the target character in the original word $w$ was the $i$th character from $C$, given that the character in the same position in $w^*$ is the $j$th character from $C$. By Bayes' rule this is equal to:
\[
\Pr[ \mathsf{C}_i \mid \mathsf{C}^*_j ] = \frac{\Pr[ \mathsf{C^*}_j \mid \mathsf{C}_i] \Pr[\mathsf{C}_i]} {\Pr[\mathsf{C}^*_j]}
\]
From the description of $k$-RR 
\[
\Pr[ \mathsf{C}^*_j \mid \mathsf{C}_j] = 1 \cdot (1 - \gamma) + 0 \cdot \gamma = 1 - \gamma,
\]
and for $i \neq j$
\[
\Pr[ \mathsf{C}^*_j \mid \mathsf{C}_i] = 0 \cdot (1 - \gamma) + \frac{1}{k-1} \cdot \gamma = \frac{\gamma}{k-1},
\]
and thus, 
\begin{align*}
    \Pr[ \mathsf{C}^*_j \mid \mathsf{C}_i] &= (1 - \gamma) \delta_{i, j} + \frac{\gamma}{k-1} ( 1 - \delta_{i, j}) \\
    &= \frac{\gamma + ((1 - \gamma)(k-1) - \gamma ) \delta_{i, j}}{k - 1},
\end{align*}
where $\delta_{i, j} = 1$ if $i = j$ and $0$ otherwise. From this, we can expand:
\begin{align*}
\Pr[\mathsf{C}^*_j] &= \sum_{\ell = 1}^n \Pr[ \mathsf{C}^*_j \mid \mathsf{C}_\ell] \Pr[\mathsf{C}_\ell] \\
&= \left(\sum_{\ell = 1, \ell \neq j}^n \frac{\gamma}{k-1} \Pr[\mathsf{C}_\ell] \right) + \left(1 - \gamma \right) \Pr[\mathsf{C}_j] \\
&= \frac{\gamma}{k-1} (1 - \Pr[\mathsf{C}_j]) + \left(1 - \gamma  \right) \Pr[\mathsf{C}_j] \\
&= \frac{\gamma}{k-1} + \left(1 - \gamma - \frac{\gamma}{k - 1}\right)  \Pr[\mathsf{C}_j] \\
&= \frac{\gamma + ((1 - \gamma)(k-1) - \gamma ) \Pr[\mathsf{C}_j]}{k-1}
\end{align*}
Combining these results, and simplifying we get:
\begin{equation}
\label{eq:main}
    \Pr[ \mathsf{C}_i \mid \mathsf{C}^*_j ] = \frac{(\gamma + ((1 - \gamma)(k-1) - \gamma  ) \delta_{i, j})\Pr[\mathsf{C}_i]}{\gamma + ((1 - \gamma)(k-1) - \gamma ) \Pr[\mathsf{C}_j]}.
\end{equation}
In general, the probabilities $\Pr[\mathsf{C}_i]$ for all $i \in |C|$ are difficult to estimate as they depend on a number of factors including the context of the text, other words surrounding a given word, and characters in the same word. On the other hand, if a word is constructed by sampling each character uniformly at random from $C$ then we can substitute $\Pr[\mathsf{C}_i] = 1/k$ for all $i$ in Eq.~\eqref{eq:main} to obtain the simplified equation:   
\begin{equation}
\label{eq:random-w-char}
\Pr[ \mathsf{C}_i \mid \mathsf{C}^*_j, w \text{ random} ] = p_{i, j, \gamma} =  \begin{cases}
    1 - \gamma &\text{ if } j = i,\\
    \frac{\gamma}{k-1} &\text{ if } j \neq i
\end{cases},
\end{equation}
and because $w$ is constructed by selecting each character uniformly at random, we get:
\begin{equation}
\label{eq:random-w}
\Pr[ \mathsf{W} \mid \mathsf{W}^*, w \text{ random} ] = p_{i,j,\gamma}^{|w|}.
\end{equation}

This gives us two competing hypothesis for each character of the original word $w$: one that says that each character of $w$ is the same as the character of $w^*$, and the other which says that each character is different. Given the length of the word $w$ we can plot the \emph{log likelihood ratio} of these two probabilities across different values of $\epsilon$ (and hence $\gamma$) as
\begin{align*}
    \ln\left( \frac{p_{j,j,\gamma}^{|w|}}{\left(\sum_{i=1, i \neq j}^{|C|} p_{i,j,\gamma}\right)^{|w|}} \right) &= \ln\left( \frac{p_{j,j,\gamma}^{|w|}}{\left(1 - p_{j,j,\gamma} \right)^{|w|}} \right)\\
&= |w| \left(\ln (1 - \gamma) - \ln( \gamma) \right)
\end{align*}

\begin{figure}[h]
    \centering
    \begin{subfigure}{0.45\textwidth}
    \includegraphics[width=\textwidth]{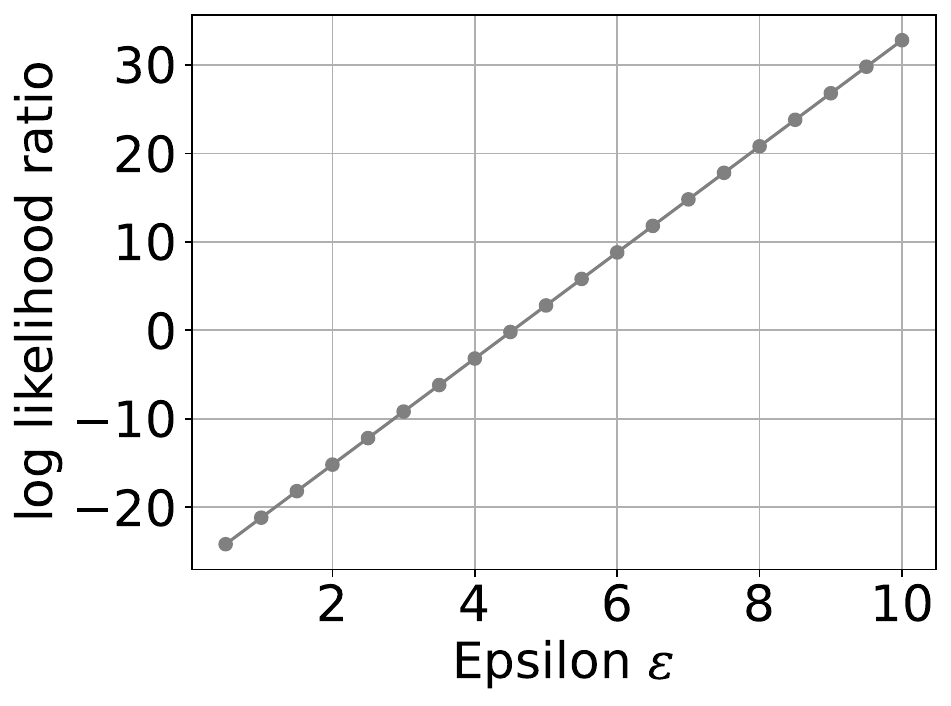}
    \captionsetup{justification=centerlast}
    \caption{Log-likelihood Ratio}
    \end{subfigure}
    \hfill
    \begin{subfigure}{0.45\textwidth}
    \includegraphics[width=\textwidth]{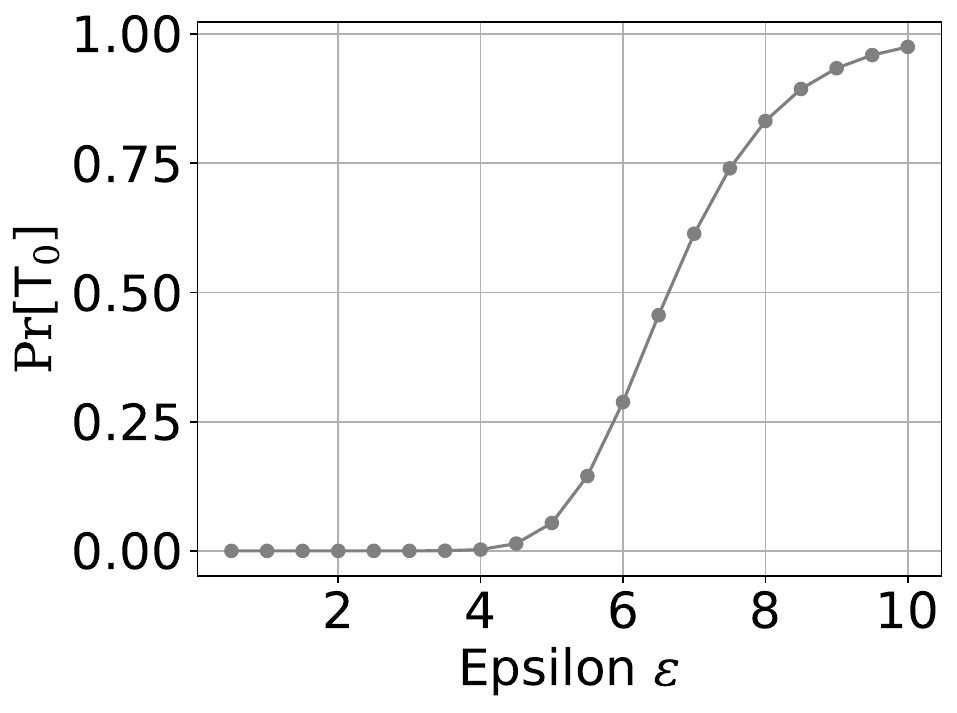}
    \captionsetup{justification=centerlast}
    \caption{Probability from Eq.~\eqref{eq:random-w}}
    \label{subfig:llr-prob}
    \end{subfigure}%
    \captionsetup{justification=centerlast}
    \caption{The log-likelihood ratio and the probability that the original word $w$ is the same as the perturbed word $w^*$ for a word of 6 characters constructed randomly across different values of $\epsilon$.}
    \label{fig:llr}
\end{figure}

In Figure~\ref{fig:llr} we plot the log-likelihood ratio and the probability in Eq.~\eqref{eq:random-w} with $i = j$ and $|w| = 6$. We can see that past $\epsilon = 4$, the log-likelihood ratio is positive, meaning that any algorithm who has been tasked to guess the word $w$ behind $w^*$, can use the hypothesis that all characters were not perturbed by the $k$-RR algorithm. The probability of correctly guessing the target word improves just after $\epsilon = 4$. 

In general, some characters in the guessed word $w'$ might be the same as $w$, while others might be different. Thus, the probability that the guessed word has up to $0 \leq \ell \leq |w|$ characters different from $w$ is given as:
\begin{align}
\label{eq:random-w-binom}
\Pr[ \mathsf{W}_\ell \mid \mathsf{W}^*, w \text{ random} ] &= \sum_{m = 0}^\ell \binom{|w|}{m} p_{j,j,\gamma}^{|w| - m} ( 1 - p_{j,j,\gamma})^m \nonumber\\
&= \sum_{m = 0}^\ell \binom{|w|}{m} \left( 1 - \gamma \right)^{|w| - m} \gamma^m \nonumber\\
&= q_{\ell, |w|, \gamma}
\end{align}

Note that 
\[
\Pr[ \mathsf{W}_0 \mid \mathsf{W}^*, w \text{ random} ] = \Pr[ \mathsf{W} \mid \mathsf{W}^*, w \text{ random} ], 
\]
implies that $q_{0, |w|, \gamma} = p_{j, j, \gamma}^{|w|}$, i.e., the probability that the guessed word is exactly the same as the original word. To calculate the probability of guessing all the words in the prompt $T$, let us define the event $\mathsf{T}_\alpha$, which says that any guessed word $w'$ is at a Hamming distance $d(w, w') \leq \alpha |w|$ from the corresponding word $w \in T$, where $\alpha \in [0, 1]$. Let $\ell(i) = \lfloor \alpha i \rfloor$. Let $T(i)$ be the number of words in $T$ of length $i$. Then,
\begin{equation}
\label{eq:total}
    \Pr[\mathsf{T}_\alpha] = \frac{1}{|T|} \sum_{i = 1}^\infty q_{\ell(i), i, \gamma} T(i) 
\end{equation}
Thus, we can estimate the performance of the algorithm by checking what percentage (fracton) of words are reconstructed by it at a relative Hamming distance $\alpha$ from the actual words. If the resulting percentage is close to the probability in Eq~\eqref{eq:total} then we can say that the algorithm is only succeeding by chance. We are particularly interested when $\alpha = 0$, i.e., reconstructing the original word, and therefore the probability $\Pr[\mathsf{T}_0]$. This estimates the percentage (fraction) of words completely reconstructed by the algorithm by chance. In the following, we use this as the \emph{baseline}. As we show later, the percentage of sensitive words reconstructed by the LLM closely follows the curve for $\Pr[\mathsf{T}_0]$. This implies that the LLM is performing close to random guess in reconstructing sensitive words. A plot for $\Pr[\mathsf{T}_0]$ where all words in $T$ are of length $6$ is shown in Figure~\ref{subfig:llr-prob}.

\subsection{Prompt Restoration using LLMs}
\label{subsec:prompt-restore}

After the application of $k$-RR , 
the resultant text $T^*$ contains random substitutions that are only partially coherent. 
To minimize utility loss and enhance readability of the prompt we adopt an LLM based approach for text enhancement by restoration of distorted words. LLMs trained on extensive corpus of text data can identify and correct irregularities in text by leveraging contextual awareness and  training data diversity, unlike simple rule-based or dictionary-based correction tools. 
In our threat model, the restoration is performed by a \emph{remote} LLM $\mathcal{R}$. We therefore send $T^*$ to $\mathcal{R}$ and instruct it to first restore the text and then perform the downstream task.

For clarity and reproducibility, we present the unified  \emph{system prompt} used in our implementation below. During our analysis, we additionally separated restoration and summarization into two sequential prompts (restore $\rightarrow$ summarize) to inspect the restored text independently and to measure the effect of restoration without conflating it with the summarization stage.

\begin{mybox}[]{System Prompt}
You are a text restoration and summarization assistant.
First, correct only the errors introduced by distortion/noise. Do not make any unnecessary changes. Preserve the original wording, punctuation, capitalization, and formatting as much as possible.
Second, create a concise and accurate summary of the restored text. Focus on the main ideas and key details, and avoid unnecessary details. Do not add opinions or any prefacing.
\end{mybox}

\section{Experimental Setup}
In this section we give details of our experimental setup and evaluation metrics.

\subsection{Datasets and Sensitive Terms}
\label{subsec:datasets}
Our evaluation is done on two datasets from different domains.
\begin{itemize}
    \item \emph{Health dataset:} This is the i2b2/UTHealth dataset \citep{Stubbs2015} which contains 1,304 longitudinal medical records describing 296 patients. This consists of multiple categories of protected health information (PHI) including names, locations, emails, phone numbers and medical record numbers (MRNs) containing on average 22.14 instances of PHI per record.
    \item \emph{Enron dataset:} The Enron dataset \citep{Enron} contains about 600,000 emails belonging to 158 employees of Enron Corporation containing personal information (PI) such as names, work addresses, phone numbers and emails. We randomly sample a subset of 10,000 records from the dataset, which averages 8.02 instances PI per record.
\end{itemize}
Note that our selection of what is considered personal information is natural, in the sense that these are the pieces of information that would most likely be considered as sensitive in these types of datasets. In particular, we have not excluded/included any category of sensitive information that would likely give an advantage to our proposed scheme. 

\subsection{LLMs} 
We experiment with two models as the Remote LLM: (1) \texttt{GPT-4o mini}\footnote{\url{https://platform.openai.com/docs/models\#gpt-4o-mini}}, a lightweight variant of \texttt{GPT-4o} \citep{openai2024gpt4ocard}, optimized for low-latency and cost-efficient text processing; and (2) \texttt{Llama-3.1 8B Instruct} \citep{grattafiori2024llama3herdmodels}\footnote{\url{https://huggingface.co/meta-llama/Llama-3.1-8B-Instruct}}, an open-source 8B-parameter model capable of handling complex language tasks such as text correction and summarization. 

\subsection{Privacy and Utility Metrics}
\label{subsec:metrics}
We use the following metrics to evaluate the utility and privacy of our experimental results.

\paragraph{\textbf{Semantic Similarity:}} To measure how well the meaning of original text is preserved after noise injection and prompt restoration, we use cosine similarity through a sentence transformer model \texttt{all-distilroberta-v1}.\footnote{\url{https://huggingface.co/sentence-transformers/all-distilroberta-v1}} This model is designed to map sentences and paragraphs into a 768-dimensional dense vector space, to facilitate tasks like clustering and semantic search. It is built upon the distilled version of RoBERTa~\citep{Roberta} architecture and was fine-tuned on over a billion sentence pairs from diverse datasets. To compare the similarity of two sentences, they are first converted into sentence embeddings using this model and then cosine similarity is calculated over these embeddings. We call the final similarity score, \emph{semantic similarity},  which is cosine similarity  converted to a percentage between 0 and 100.

\paragraph{\textbf{Reconstruction Rate:}} The differential privacy parameter $\epsilon$ dictates the probability with which each character is perturbed in a word. However, we need another privacy metric to assess the difficulty in reconstructing a complete word. For this we use the \emph{reconstruction rate}, which is the percentage of terms reconstructed by an LLM. This can be used to quantify the number of sensitive or non-sensitive words that are successfully reconstructed through prompt enhancement. Note that for the latter category of words, the reconstruction rate serves as a utility metric. In other words, we want high reconstruction of non-sensitive words, and low reconstruction of sensitive words.  

\subsection{What Qualifies as a Rare Word?}
\label{subsec:Reconstructed_Words}
Our approach is motivated by the observations  that non-sensitive and sensitive words exhibit distinct statistical and linguistic properties, which directly influence how well an LLM can reconstruct distorted text. LLMs are widely used for various NLP tasks, yet they often struggle to handle typos and misspellings of rare or unseen words, leading to inaccurate outputs. According to Fox et al. \citep{Fox2022} end to end models struggle with words that are rarely or never seen during model training. Likewise, Zhu et al. \citep{PromptRobust} demonstrated that lower frequency words are inherently less robust to even minor perturbations like typos, synonym swaps or restructured sentences. The study by Schick et al. \citep{Schick2020} also reveals that even models like BERT struggle to understand rare words as they rely on sub-word based tokenization which does not always assign dedicated embeddings to infrequent words. This results in rare words not being represented in the model's learned vocabulary leading to poor semantic understanding.
Based on this discussion, we hypothesize that:

\begin{itemize}
    \item Sensitive words are typically high-entropy and rare, meaning they are less frequently found in training corpora.
    \item Non-sensitive words are generally low-entropy and common, that are more frequently seen by the model during training.
    \item LLMs reconstruct non-sensitive words better than sensitive words due to their prior exposure to common vocabulary and sentence structures compared to rare sensitive words.
\end{itemize}

\section{Results and Analysis}
We begin this section with the overall performance of our method in sanitizing sensitive words. 

\subsection{Word Reconstruction}
We begin by running our method on the i2b2/UTHealth and Enron datasets with varying levels of $\epsilon$, and evaluating privacy and utility of the resulting prompt using the metrics described in the previous section. More specifically, privacy is measured through the reconstruction rate of sensitive words, while utility is quantified through the reconstruction rate of non-sensitive words as well as the semantic similarity between the original summary and the restored summary.

\begin{figure*}[h!]
    \centering
    \begin{subfigure}{0.45\textwidth}
        \includegraphics[width=\textwidth]{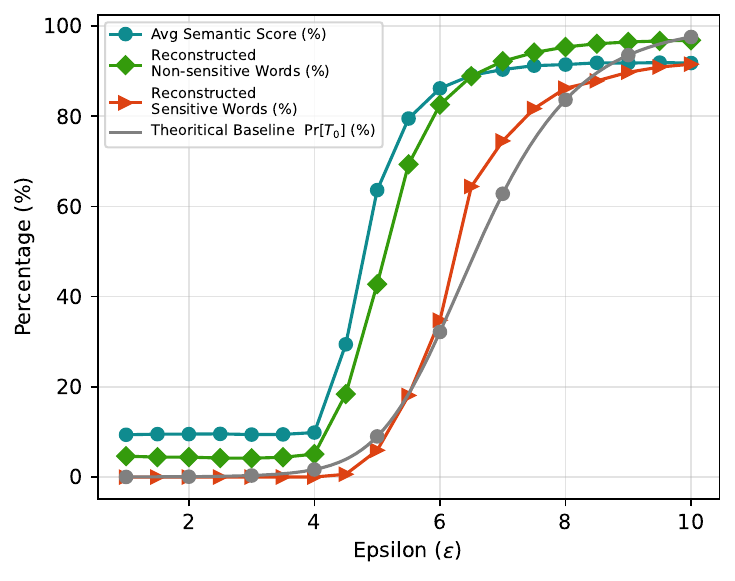}
        \captionsetup{justification=centerlast}
        \caption{Health dataset (GPT-4o mini)}
        \label{fig:Health_GPT_Reconstruction}
    \end{subfigure}
    \begin{subfigure}{0.45\textwidth}
        \includegraphics[width=\textwidth]{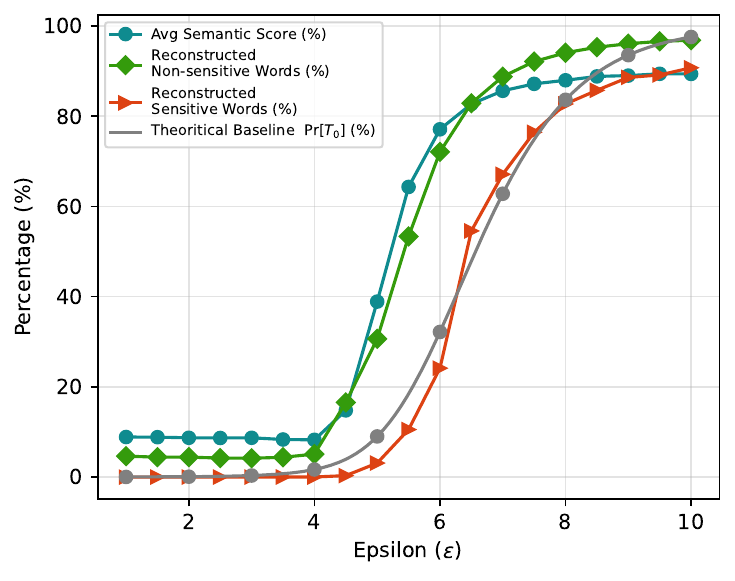}
        \captionsetup{justification=centerlast}
        \caption{Health dataset (Llama-3.1 8B)}
        \label{fig:Health_Llama_Reconstruction}
    \end{subfigure}

    \begin{subfigure}{0.45\textwidth}
        \includegraphics[width=\textwidth]{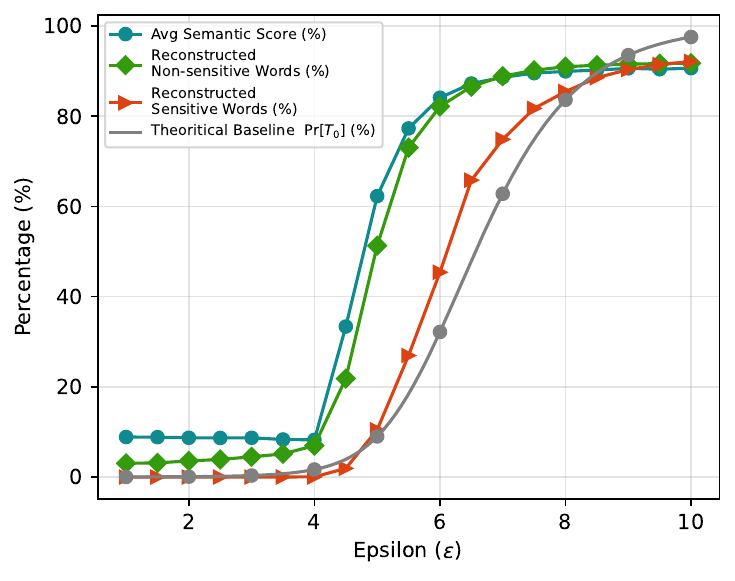}
        \captionsetup{justification=centerlast}
        \caption{Enron dataset (GPT-4o mini)}
        \label{fig:Enron_GPT_Reconstruction}
    \end{subfigure}
    \begin{subfigure}{0.45\textwidth}
        \includegraphics[width=\textwidth]{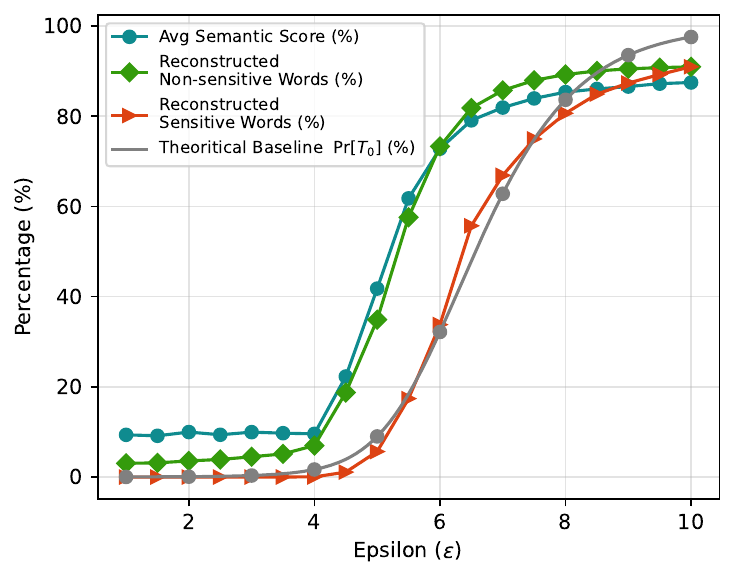}
        \captionsetup{justification=centerlast}
        \caption{Enron dataset (Llama-3.1 8B)}
        \label{fig:Enron_Llama_Reconstruction}
    \end{subfigure}
    
    \caption{Sensitive, non-sensitive terms reconstruction (\%) by remote LLMs,  theoretical reconstruction baseline $\Pr[\mathsf{T}_0]$ from Eq.~\eqref{eq:total}, and average semantic similarity (\%) vs. $\epsilon$ for the i2b2/UTHealth and Enron datasets.}
    \label{fig:i2b2_UTHealth_enron_utility_reconstruction_analysis}
\end{figure*}

Figures~\ref{fig:i2b2_UTHealth_enron_utility_reconstruction_analysis} shows the results of applying our mechanism, noise injection, followed by prompt restoration via the remote LLMs, GPT-4o mini and Llama-3.1 8B, on the i2b2/UTHealth and Enron datasets, respectively. The  theoretical baseline $\Pr[\mathsf{T}_0]$ curve is plotted through Eq.~\ref{eq:total} in Section~\ref{sec:krr-analysis}. This is the probability of reconstructing words, if the sensitive words were to be replaced by words of the same length each character of which is selected uniformly at random from the 94 printable ASCII characters. Thus, this serves as a baseline for the reconstruction rate.

At lower values of $\epsilon < 4$, none of the sensitive words are reconstructed. However, about 10\% of non-sensitive words are still being reconstructed. As $\epsilon$  increases beyond 4, reconstruction rates for both sensitive and non-sensitive words rise sharply and plateau at high values by $\epsilon = 10$; the residual shortfall is expected because our exact match metric, which penalizes even plausible generative variations. In some cases, the restoration model may output a phonetically or orthographically similar but distinct entity, such as restoring `Stevenson' as `Stephenson' or `Stafford' as `Stratford'. Similarly, occasional minor character variations such as 43 year old' versus 43-year- old' prevent exact matches despite semantic equivalence.

Our first main observation is that the reconstruction rate of the sensitive words is close to the theoretical baseline for both datasets and both LLMs. On the other hand, there is a gap between the reconstruction rate of sensitive vs non-sensitive words, especially between $4 \leq \epsilon \leq 10$. This gives credence to our hypothesis that LLMs would easily reconstruct common words, which would mostly be non-sensitive, while struggle with rare words, of which most are sensitive. 

The other important thing to notice is that the semantic similarity of the original prompt summary with the restored prompt summary is high, even if a low percentage of non-sensitive words have been reconstructed. For instance, in \autoref{fig:Health_GPT_Reconstruction}, at $\epsilon = 5.5$, the reconstruction rate of sensitive words is less than 20\%, and that of non-sensitive words is about 70\%, yet the similarity score is 80\%. Thus, this suggests that an $\epsilon$ between the ranges of $4$ to $6$ should give us a good balance between privacy and utility. We will return to the choice of $\epsilon$ in Section~\ref{subsec:epsilon}. For now, we are interested in knowing what causes sensitive words to be reconstructed at a substantially lower rate than non-sensitive words. After all, we did not use any specific criterion to choose sensitive words other than what may be PII.

\subsection{Impact of Iterative Restoration}
An interesting question to ask is whether the LLM can improve prompt restoration if it is asked to better its previous attempt. In order to check this, we  performed a second restoration with the same system prompt shown in Section~\ref{subsec:prompt-restore}. 
While the first pass ensures most of the utility is recovered with controlled privacy, the second pass helps us to evaluate whether iterative restoration increases the likelihood of sensitive information reconstruction. Since a second restoration is resource intensive, we selectively performed this experiment on i2b2/UTHealth dataset (highest PII instances per record) using GPT-4o mini, which was identified as the best performing LLM in the first phase of reconstruction. As shown in Table~\ref{tab:Secondreconstruction}, the restoration percentages indicate that applying an additional LLM restoration step does not lead to a significant increase in the overall reconstruction percentage of sensitive terms. While this second reconstruction introduces new reconstructed terms, it was observed that out of the newly reconstructed total terms 98.72\% are non-sensitive words while only 1.28\% are sensitive words. This further supports our hypothesis that LLMs still struggle to reconstruct rare, sensitive words.

\begin{table}[h] 
    \centering
    \caption{Comparison of First  and Second Restorations for Sensitive and Non-Sensitive Words (i2b2/UTHealth dataset - GPT-4o mini) }
    \label{tab:Secondreconstruction}
    \resizebox{\columnwidth}{!}{%
    \begin{tabular}{|c|c|c|c|c|}
        \hline
        \multirow{3}{*}{Epsilon ($\epsilon$)} & \multicolumn{2}{c|}{Non-sensitive Words} & \multicolumn{2}{c|}{Sensitive Words} \\
        \cline{2-5}
        & First & Second & First & Second \\
        & Restoration (\%) & Restoration (\%) & Restoration (\%) & Restoration (\%) \\
        \hline\hline

        1.0  & 4.54 & 4.54 & 0.00 & 0.00 \\
        1.5  & 4.32 & 4.32 & 0.00 & 0.00 \\
        2.0  & 4.29 & 4.29 & 0.00 & 0.00 \\
        2.5  & 4.04 & 4.05 & 0.00 & 0.00 \\
        3.0  & 4.01 & 4.01 & 0.00 & 0.00 \\
        3.5  & 4.12 & 4.12 & 0.01 & 0.01 \\
        4.0  & 4.74 & 4.77 & 0.01 & 0.01 \\
        4.5  & 17.88 & 18.05 & 0.62 & 0.71 \\
        5.0  & 42.86 & 43.70 & 5.96 & 6.53 \\
        5.5  & 69.98 & 70.68 & 17.85 & 19.30 \\
        6.0  & 83.07 & 83.30 & 33.94 & 34.77 \\
        6.5  & 89.22 & 89.31 & 63.96 & 64.61 \\
        7.0  & 92.45 & 92.49 & 74.14 & 74.80 \\
        7.5  & 94.27 & 94.31 & 80.79 & 81.01 \\
        8.0  & 95.43 & 95.46 & 85.27 & 86.02 \\
        8.5  & 96.12 & 96.13 & 87.22 & 87.46 \\
        9.0  & 96.50 & 96.50 & 89.48 & 90.14 \\
        9.5  & 96.72 & 96.72 & 90.48 & 90.81 \\
        10.0 & 96.86 & 96.86 & 91.13 & 91.13 \\

        \hline
    \end{tabular}%
    }
\end{table}

\subsection{Reconstruction of Sensitive Entities}

The analysis in the previous section shows that the utility of the downstream task can be maintained while removing most of the sensitive terms. However, some sensitive terms are still reconstructed. This does not always mean the full sensitive \emph{entity} is recovered. Many sensitive items are entities, not single words, e.g., names and locations often have multiple parts. A model may restore one part but not the full entity. This can make leakage look higher if we only count reconstructed words.

We study this effect using the i2b2/UTHealth dataset. We use i2b2 because it has gold annotations for names and location. The Enron dataset does not provide the same type of entity labels. For names, we report three measures: total name reconstruction, single-name reconstruction, and full-name reconstruction. Full-name reconstruction means that every part of a multi-word name must be restored. For location, we follow the dataset labels. Location information is split into components such as street, city/state, and facility names (e.g.,  hospital/clinic).
We distinguish between complete address reconstruction, a strict record level metric requiring the simultaneous recovery of all address fields (e.g., Street + City + State) for a single record, and total location reconstruction, which measures the aggregate recovery of individual location entities independently. Since many records in the dataset do not contain full postal addresses, the strict complete address metric is inapplicable; therefore, we report total location reconstruction alongside the component level breakdown.

\begin{figure}[h!]
    \centering
    \begin{subfigure}{0.45\textwidth}
        \includegraphics[width=\textwidth]{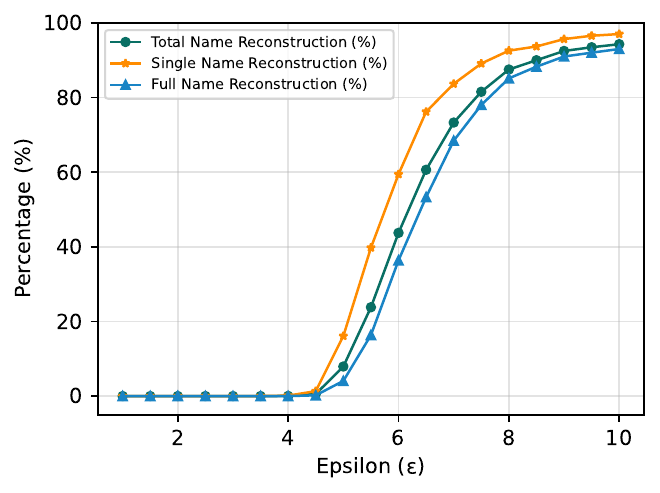}
        \captionsetup{justification=centerlast}
        \caption{GPT-4o mini}
        \label{subfig:Health_GPT_Names}
    \end{subfigure}
    \begin{subfigure}{0.45\textwidth}
        \includegraphics[width=\textwidth]{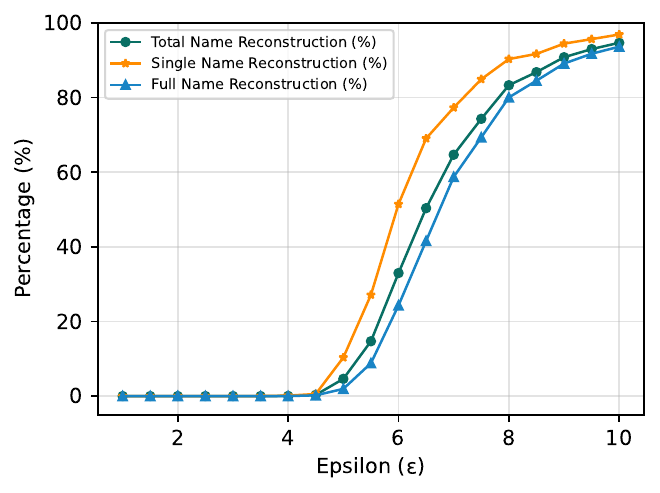}
        \captionsetup{justification=centerlast}
        \caption{Llama-3.1 8B}
        \label{subfig:Health_Llama_Names}
    \end{subfigure}    \captionsetup{justification=centerlast}

    \caption{Analysis of name reconstruction on the i2b2/UTHealth dataset across different $\epsilon$ values.}
    \label{fig:name_reconstruct}
\end{figure}

\begin{figure}[h!]
    \centering
    
    \begin{subfigure}{0.45\textwidth}
        \includegraphics[width=\textwidth]{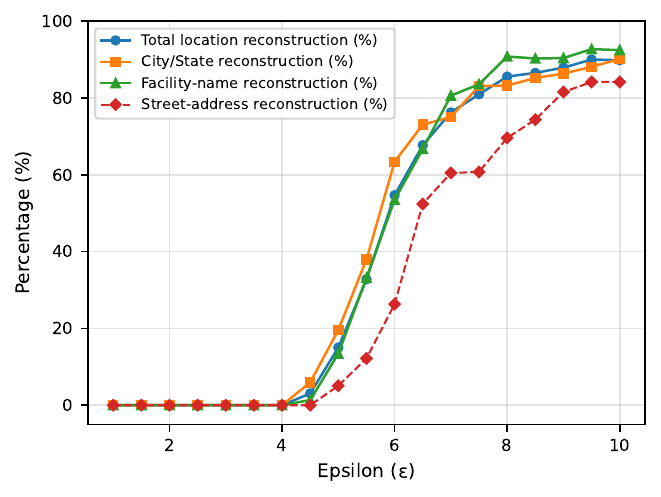}
        \captionsetup{justification=centerlast}
        \caption{GPT-4o mini}
        \label{subfig:Health(GPT)_Locations}
    \end{subfigure}    \captionsetup{justification=centerlast}
    \begin{subfigure}{0.45\textwidth}
        \includegraphics[width=\textwidth]{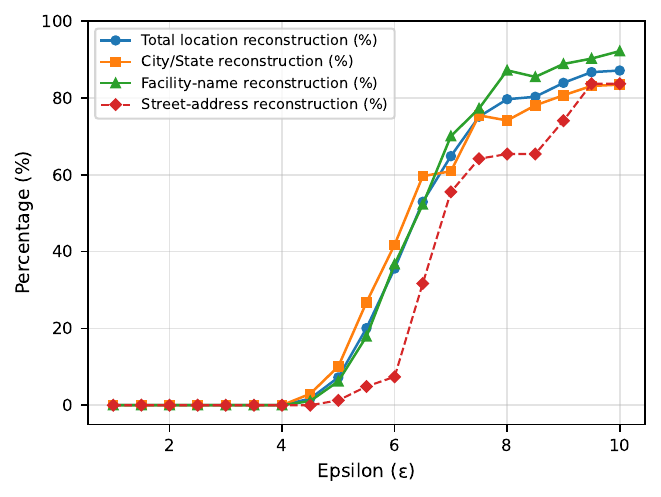}
        \captionsetup{justification=centerlast}
        \caption{Llama-3.1 8B}
        \label{subfig:HealthLlama_Locations}
    \end{subfigure}    \captionsetup{justification=centerlast}
    
    \caption{Analysis of location reconstruction on the i2b2/UTHealth dataset across different $\epsilon$ values.}
    \label{fig:location_reconstruct}
\end{figure}

Figures~\ref{fig:name_reconstruct} and \ref{fig:location_reconstruct} show the trend for both GPT-4o mini and Llama-3.1 8B for the name and location entities, respectively. For names, full-name reconstruction is lower than single-name reconstruction at all $\epsilon$ values. This means full multi-word names are harder to restore. For location, street reconstruction is lower than city/state reconstruction. Street strings are more specific and less predictable. They often include a number and an uncommon name. Examples include “50 Indian Rock Road” and “558 Paxton Terrace”. City and state terms come from a smaller and more common set, so they are easier to recover. We also see that facility names are reconstructed more than street names. Facility names are usually more structured than street strings. They often look like “X Hospital” or “Y Memorial Hospital”, as in “Pennsylvania Hospital” and “Superior Memorial Hospital”. These patterns make facility names easier to restore from partial cues.

These results strengthen the privacy claim in two ways. First, complete reconstruction of multi-word name entities is much less likely than reconstruction of single-word names. Second, for location, street-level details are harder to recover than city/state and facility names, especially at lower-to-mid $\epsilon$ values.

\subsection{Recommended Range of $\epsilon$} \label{subsec:epsilon}
To recommend the best $\epsilon$ based on our analysis, we revisit
\autoref{fig:i2b2_UTHealth_enron_utility_reconstruction_analysis}
which illustrates the trade-off between utility and privacy across different $\epsilon$ values for both GPT-4o mini and Llama-3.1 8B on the i2b2/UTHealth and Enron datasets.

At $\epsilon = 5.5 $ GPT-4o mini achieves high utility ($\sim 80\%$) , while sensitive term reconstruction remains low ($\sim 20-25\%$)  for both datasets, ensuring strong privacy protection without significantly compromising text utility. In contrast  Llama-3.1 8B, $\epsilon = 6.0$ provides similar utility ($\sim 75\%$)  while keeping sensitive term reconstruction around 30\%, making it a more privacy-utility preserving choice.
Thus, we recommend $\epsilon = 5.5 $ for GPT-4o mini and $\epsilon = 6.0 $  for Llama-3.1 8B as the optimal values that provide the best balance between privacy and utility.

\section{Comparison with Other Prompt Privacy Mechanisms}

We compare our character-level local differential privacy (LDP) method with two other prompt privacy mechanisms: a word-level differential privacy approach and a rule-based PII redaction baseline. For word-level DP, we adopt the approach of \citep{feyisetan2020privacy}, which perturbs each word by adding calibrated noise in a pre-trained embedding space and then decoding the noised vector back to a discrete token via nearest-neighbor replacement. For the rule-based baseline, we use Microsoft Presidio,\footnote{\url{https://microsoft.github.io/presidio/}} an open-source framework that detects and anonymizes PII. We evaluate all methods on two datasets (Health and Enron) and under two restoration settings (GPT-based and Llama-based) to capture both domain and model variability.

A key challenge is that these mechanisms cannot be compared fairly through a shared privacy budget. Although both character-level and word-level mechanisms are parameterized by $\epsilon$, the same numeric $\epsilon$ does not correspond to the same effective perturbation strength because the mechanisms operate at different granularities (characters versus words). Presidio further differs in that, as it is non-parametric: it does not expose an $\epsilon$ style parameter and instead produces a single deterministic transformation for a given input. For these reasons, a direct  $\epsilon$ comparison would be misleading.

To enable a fair comparison across such heterogeneous mechanisms, we evaluate all methods using a common privacy–utility perspective. Privacy is quantified via an attacker oriented measurement of sensitive term reconstruction rate. To align the metric so that higher values indicate stronger protection, we report the `privacy preserved score', defined as the complement of the sensitive term reconstruction rate ($1 - \text{Sensitive Term Reconstruction Rate}$). Utility is measured at the downstream task level. We compute cosine similarity between the summary generated from the original input and the summary generated from the protected input. We report both privacy preserved score and cosine similarity on a 0–100\% scale for readability. This yields a curve across $\epsilon$ for DP-based methods and a single operating point for Presidio.

To make this comparison fair, we standardize the summarization pipeline across all methods. After applying word-level noise or Presidio redaction, we use the same system instruction for the remote LLM. If the input appears degraded, it should enhance or fix it when possible before producing the summary. Because the summarization model receives the same enhancement instruction in every setting, variations in privacy and utility are attributable to the noise or redaction step, not inconsistencies in the summarization prompt.

\begin{figure*}[h!]
    \centering
    \begin{subfigure}{0.45\textwidth}
        \includegraphics[width=\textwidth]{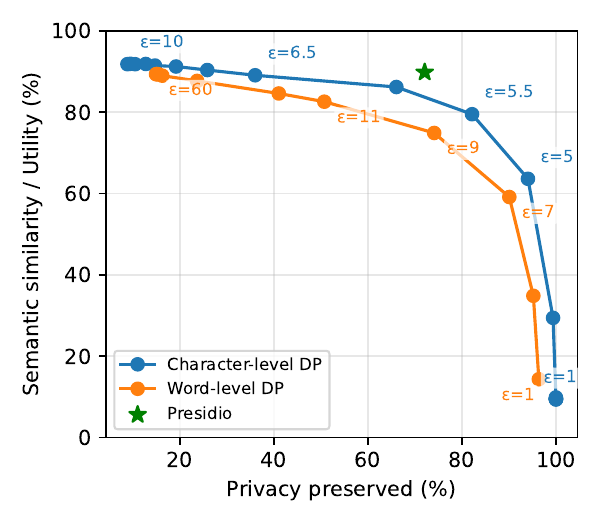}
        \captionsetup{justification=centerlast}
        \caption{Health dataset (GPT-4o mini)}
        \label{fig:Health(GPT)Reconstruction}
    \end{subfigure}
    \begin{subfigure}{0.45\textwidth}
        \includegraphics[width=\textwidth]{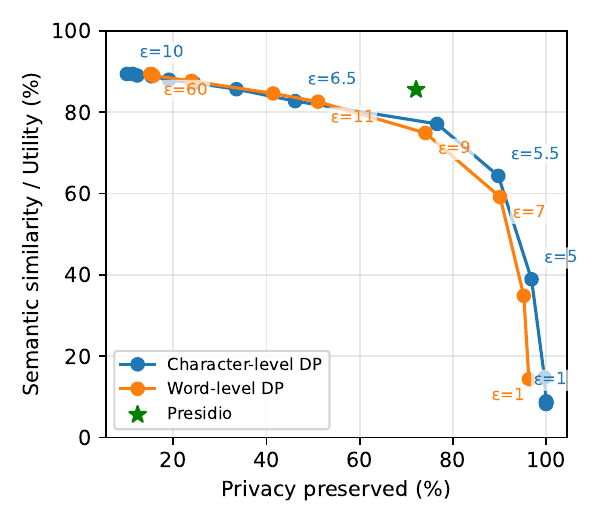}
        \captionsetup{justification=centerlast}
        \caption{Health dataset (Llama-3.1 8B)}
        \label{fig:Health(Llama)Reconstruction}
    \end{subfigure}

    \begin{subfigure}{0.45\textwidth}
        \includegraphics[width=\textwidth]{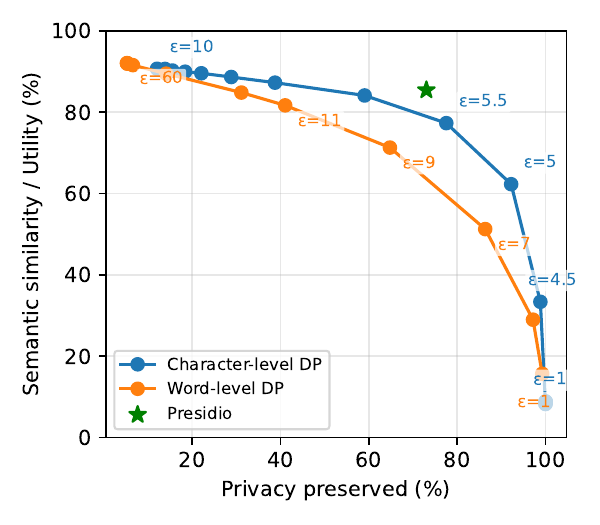}
        \captionsetup{justification=centerlast}
        \caption{Enron dataset (GPT-4o mini)}
        \label{fig:Enron(GPT)Reconstruction}
    \end{subfigure}
    \begin{subfigure}{0.45\textwidth}
        \includegraphics[width=\textwidth]{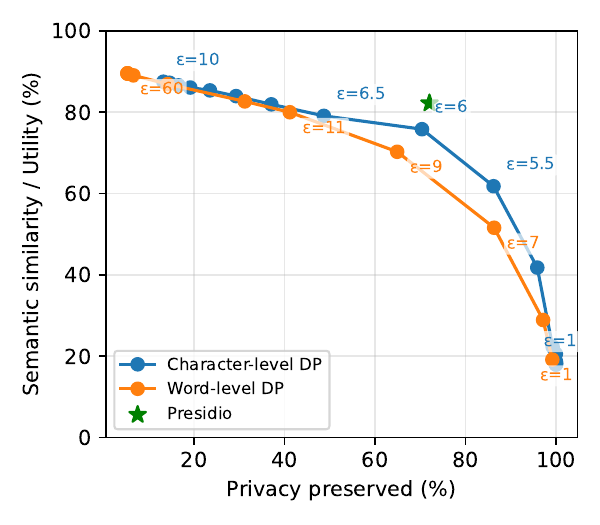}
        \captionsetup{justification=centerlast}
        \caption{Enron dataset (Llama-3.1 8B)}
        \label{fig:Enron(Llama)Reconstruction}
    \end{subfigure}
    
    \caption{Baseline Comparison}
    \label{fig:Baseline_comparison}
\end{figure*}

Overall, the baseline comparison in \autoref{fig:Baseline_comparison} indicates that our character-level LDP method is a practical option alongside word-level DP and Presidio. When compared with word-level DP, character-level perturbation tends to retain summarization quality better at similar levels of privacy protection. This suggests that character-level noise introduces more local distortions that the restoration model can often repair, without necessarily reconstructing sensitive terms exactly. By contrast, word-level perturbation replaces whole tokens. This can introduce larger semantic shifts, which are harder to undo during enhancement, and can lead to a less favorable privacy-utility trade-off.

On the other hand, Presidio typically achieves high utility at its fixed setting. This is expected because it does not add random noise and often leaves non-sensitive text unchanged. However, Presidio provides only a single operating point and therefore offers limited flexibility when privacy requirements differ across contexts. In practice, deployments often need to adjust protection depending on trust assumptions and risk. For example, summarizing a patient note using a third-party LLM may require a stricter privacy setting than an internal workflow operating in a more controlled environment. DP based mechanisms naturally support this context-dependent deployment by allowing a range of operating points: stronger protection (smaller $\epsilon$) can be chosen when risk is high, while lighter perturbation (larger $\epsilon$) can be used when utility is more critical. In comparison, both DP mechanisms support multiple operating points across the privacy–utility spectrum. Within that spectrum, our results suggest that character-level perturbation provides a more preferable set of operating points than the word-level alternative.

Another difference is that Presidio’s protection is limited to what it can actually detect. When Presidio detects an entity, it can redact it effectively, but if an identifier is not detected due to an uncommon format, a domain-specific code, or a misspelling, it may be left unchanged and receive no protection. This matters in realistic text, where sensitive information is not always neatly formatted as standard PII categories. For example, a clinical note may include an internal identifier such as `MRN 00-34-81-92,' or a contact reference such as `j.holmes ext 74219,' which can be identifying in context but may not match a predefined recognizer. In contrast, our character-level LDP perturbation applies independently of detection, so such identifiers are still likely to be distorted and therefore harder to reconstruct exactly.

\section{Discussion}

Our privacy guarantee is based on the fact that the $k$-RR algorithm perturbs characters in sensitive words to such an extent that even an LLM is not able to recover the original word. In essence, we use the LLM as a proxy for what is recoverable, in the sense that any other algorithm, or even human, will not be able to do significantly better. Since LLMs are state-of-the-art in such textual tasks, we believe that this assumption is well-grounded.

In our setting, many sensitive terms are high-entropy strings such as uncommon names or identifiers, and their exact spelling is only weakly constrained by surrounding context. By contrast, common words are often recoverable because syntax, word length, and context sharply restrict the candidate set even when several characters are corrupted; for example, ``Pje2\$e c4pq a d\&ctBr'' is still readily interpreted as ``Please call a doctor.'' Proper names and identifiers do not benefit from the same constraints. Suppose the true name is ``Chervy,'' but after perturbation it appears as ``Ch\%r\$y'' in the sentence ``Ch\%r\$y is a residential patient.'' Even if it is clear that the token is a name and the rest of the sentence is intact, many alternatives remain plausible (e.g., ``Cherry,'' ``Charly,'' ``Sherry''), and there is typically no contextual signal strong enough to select the correct one. Once character-level perturbation disrupts these distinguishing spelling cues, exact recovery becomes difficult without auxiliary information, regardless of whether the restorer is a human or a model.

Auxiliary knowledge can change this picture by shrinking the search space. If an adversary already has access to a patient list, employee directory, or a small candidate set of identifiers, they could attempt a constrained guess and check strategy. However, this shifts the threat model from prompt only recovery to a linkage attack that depends on external data. Here in our evaluation, we treat strong modern LLMs as a reasonable proxy for the best available text only restorer, so failure to recover high entropy tokens suggests that the remaining signal is insufficient for reliable exact reconstruction.

Our definition of private terms is guided by the sensitive categories present in our evaluated datasets, such as names and locations. In other domains, sensitive content can also include business or technical identifiers like IP addresses, software versions, internal ticket IDs, or configuration strings, where leakage can have direct security consequences. A practical implication is that the privacy–utility trade-off depends on the restoration model: the same epsilon can lead to different reconstruction behavior under different LLM restorers. This suggests that epsilon should be calibrated for the specific restoration model used in deployment. A conservative strategy is to calibrate epsilon against a strong restoration model as a worst case restorer and then deploy that setting, since weaker models are unlikely to reconstruct more than the strongest model under the same perturbation. In practice, this points to a lightweight empirical calibration step to choose epsilon for a given dataset and model configuration.

We also examined whether individual word-level characteristics, such as word frequency (as a proxy for training exposure), explain the gap between sensitive and non-sensitive reconstruction. We did not find a meaningful or consistent association, suggesting that these characteristics are not the main differentiator. Instead, reconstruction appears to be driven more by contextual constraints: common words are often recoverable because the surrounding text narrows the candidate space, whereas sensitive terms typically remain underconstrained even when the rest of the sentence is intact.

\section{Conclusion}
In this paper, we introduce an approach that applies local differential privacy at the character level to sanitize text data given as input prompts to large language models . We experimentally evaluate our approach on two datasets of emails and clinical notes, with Llama 3.1 8B and GPT-4o mini models. Our findings indicate that with the application of our technique, sensitive PII is reconstructed at a rate nearly equivalent to that of random words, while non-sensitive words are recovered at much higher rates, indicating strong privacy protection with good utility preservation. 
We also compared our method against a word-level differentially private mechanism and an NER-based sanitization approach under a unified privacy–utility evaluation framework.
Our technique is agnostic of the underlying text, and serves as a useful tool to enhance prompt privacy, while maintaining prompt utility without the pitfalls of supervisory models that need to identify and sanitize private terms.



\bibliography{References}
\bibliographystyle{abbrvnat}

\appendix


\section{$k$-Randomized Response is $\epsilon$-DP}
\label{app:krr-proof}
Consider two different inputs $x, x' \in \mathcal{X}$. Fix an outcome $y$ in the (discrete) range  $\mathcal{Y}$ of $\mathcal{M}$. Note that in the $k$-RR algorithm $\mathcal{X} = \mathcal{Y} = C$, with $|C| = k$. Assume first that $y \neq x, x'$. Then
\begin{equation*}
    \Pr[ \mathcal{M}(x) = y ] = \frac{\gamma}{k-1} = \Pr[ \mathcal{M}(x') = y ] = e^0 \Pr[ \mathcal{M}(x') = y ].
\end{equation*}
 Now assume that $y = x$, then
\begin{equation*}
    \Pr[ \mathcal{M}(x) = y ] = 1 - \gamma,
\end{equation*}
and 
\begin{equation*}
    \Pr[ \mathcal{M}(x') = y ] = \frac{\gamma}{k-1},
\end{equation*}
We see that the ratio:
\begin{align*}
    \frac{\Pr[ \mathcal{M}(x) = y ]}{\Pr[ \mathcal{M}(x') = y ]} &= (k-1)\frac{1 - \gamma}{\gamma}\\
                &= \frac{k-1}{\gamma} - k + 1 \\
                &= (k-1) \frac{e^\epsilon + k - 1}{k -1} - k + 1\\
                &= e^\epsilon,
\end{align*}
where we have used the value of $\gamma$ from Algorithm~\ref{algo:krr}. From the above:
\begin{align*}
    \frac{\Pr[ \mathcal{M}(x') = y ]}{\Pr[ \mathcal{M}(x) = y ]} &= e^{-\epsilon} \leq e^{\epsilon}.
\end{align*}
The case with $y = x'$ is the same by symmetry. Therefore, for all $x, x' \in \mathcal{X}$ and all $y \in \mathcal{Y}$ we see that 
\[
\Pr[ \mathcal{M}(x) = y ] \leq e^\epsilon \Pr[ \mathcal{M}(x') = y ].
\]

\end{document}